\documentclass[intlimits,twoside,a4paper]{article}

\usepackage{graphicx}
\usepackage{amssymb,amsmath}
\usepackage[T2A]{fontenc}
\usepackage[cp1251]{inputenc}
\usepackage{cmpj}

\issue{2011}{14}{1}{13603}

\doinumber{10.5488/CMP.14.13603}

\title[The liquid-vapor interface of the restricted primitive model]
{The liquid-vapor interface of the restricted primitive model of ionic fluids
from a density functional approach}

\author[A.~Patrykiejew, S. Soko\l owski, O. Pizio]{A. Patrykiejew\refaddr{label1},
S. Soko\l owski\refaddr{label1},
O. Pizio\refaddr{label2}\thanks{E-mail: pizio@servidor.unam.mx}}

\authorcopyright{A.~Patrykiejew, S. Soko\l owski, O. Pizio}

\addresses{
\addr{label1}Department for the Modelling of Physico-Chemical Processes,
Maria Curie-Sk{\l}odowska University, \\20--031 Lublin, Poland,
\addr{label2}Instituto de Qu\'{i}mica de la UNAM, Coyoac\'{a}n 04510, M\'{e}xico, D.F.,
Mexico}

\date{Received September 13, 2010, in final form November 23, 2010}
\begin{document}
\maketitle

\begin{abstract}
We investigate the liquid-vapor interface of the restricted primitive model for
an ionic fluid using a density functional approach.
 The applied theory includes the
electrostatic contribution to the free energy functional arising from the bulk energy
equation of state and
the mean spherical  approximation  for a restricted
primitive model, as well as the associative contribution, due to the
formation of pairs of  ions. We compare the density profiles and the values
of the surface tension with previous theoretical approaches.

\pacs 68.08.-p, 68.43.Fg,  82.35.Gh, 68.43.-h
\end{abstract}

\section{Introduction}

According to the simplest model of ionic fluids, called the restricted primitive model (RPM), the
fluid consists of charged hard spheres (at total density $\rho$) of equal diameter,
$d$, half of which carry a charge of $+q$, while the other half carry a charge of
$-q$. These spheres are immersed in a dielectric medium of dielectric
constant $\epsilon$. Reduced thermodynamic quantities, temperature
and density, appropriate to this model are defined
as follows $T^*=kT\epsilon d/q^2$ and $\rho^*=\rho d^3$.

Perhaps, the first successful
 statistical-mechanical description
of this model was proposed by Debye and H\"uckel~\cite{DH} about
ninety years ago. The Debye-H\"uckel approach predicted the
existence of the first-order transition for RPM and was
subsequently applied in several works, e.g., to describe phase
behavior of three-component ionic fluids~\cite{NNN}. More
recently,  the bulk RPM  was  studied by means of the
Ornstein-Zernike (OZ) equation supplemented by the mean spherical
approximation (MSA)~\cite{1,2,2a}, which permits to obtain an
analytical solution for the pair correlation functions.
 The MSA correlation functions  can be used to
evaluate an equation of state~\cite{WL,WL1}. In spite of quite
reasonable results obtained for the structure and thermodynamics
in a general sense, neither virial nor compressibility equations
of state permit to obtain gas-liquid  transition for the RPM,
which contradicts the computer simulations
\cite{3,4,5,6,SYM,7,new7,new8}. By contrast, the equation of state
resulting from the \textit{energy} route predicts the first-order
phase transition between an ionic vapor and a dense ionic liquid
\cite{WL,WL1}. However, there is a big discrepancy between the
values of critical temperature and critical density of MSA theory
and the critical parameters resulting from computer simulations.
Moreover, a more sophisticated generalized mean-spherical
approximation~\cite{8}, which
 was designed to reconcile thermodynamic inconsistency
of the MSA, does not lead to a better agreement
of the critical parameters with simulations as well.

Since both MSA and generalized MSA theories underestimate the
critical density and overestimate the critical temperature,
attempts were made  to include \textit{ad hoc} the effect of ion
pairing (association) into the RPM, which is most pronounced along
the vapor branch of the coexistence envelope and near the critical
point~\cite{7,9}. The inclusion of the association into the theory
improves the agreement of theoretical predictions for the critical
constants with simulations (cf. table~I of~\cite{7}). However, the
overall shape of the coexistence envelope coming from several
theoretical approaches is not well described compared to
simulation data. In particular, the liquid densities along
coexistence are poorly described for different temperatures. We
should bear in mind that the Hamiltonian of the model with
association used in the theoretical developments differs from the
genuine RPM used in simulations. Recent developments in the
modelling of different types of associations and clustering in
liquids and solutions have been comprehensively and critically
reviewed by Holovko in~\cite{hol1}.

The liquid-vapor interface is one of the simplest examples of
nonuniform systems. However, compared to the case of Lennard-Jones
fluids~\cite{LJ}, much less is known about the liquid-vapor
interface in the RPM. Theoretical work on the gas-liquid interface
for RPM was pioneered by Telo da Gama et al.~\cite{Gama}, that
used the gradient expansion method~\cite{Gradient}. The problem of
describing an ionic liquid-ionic vapor interface has been also
studied by Groh et al.~\cite{Groh}, who proposed a density
functional approach to evaluate the surface tension and the
density profiles through the interface. Their approach involved a
local density approximation for the hard-sphere part of the
free-energy functional and a nonlocal treatment of Coulombic
contributions. The latter term was evaluated approximating the
inhomogeneous pair correlation functions by their bulk
counterparts. There are significant differences between the theory
used by Groh et al.~\cite{Groh} and the density functional
approaches that were proved successful in the description of
electrical double layers at charged walls
\cite{13,17,18,19,21,21a,21b,21c,21d}. Namely, the theories of
\cite{13,17,18,19,21} treat the electrostatic part of the free
energy by means of a second-order density expansion about the
density of a reference fluid, which was taken to be the
homogeneous bulk fluid far from the surface. Although such an
approach can be appropriate for many purposes, but it is
problematical when it comes to a liquid-vapor interface, where two
bulk fluids are involved. Moreover, it is known that in the case
of Lennard-Jones fluids, the corresponding second-order expansion
with respect to a homogeneous fluid fails to account for the
liquid-vapor coexistence. More recently, the description of a
liquid-vapor interface was the subject of investigations of Weiss
and Schr\"oer~\cite{23}. Using  a square-gradient type theory they
computed the density profiles and the interfacial tension at
different temperatures using Debye-H\"uckel theory and its
extension to ion-pair formation, as well as adding the free energy
term describing correlations between ion pairs as entities and
free ions, as developed by Fisher and Levin~\cite{9}.

Any density functional theory determines thermodynamic properties of an
inhomogeneous fluid from the Helmholtz free energy $F$ and its functional
dependence on the local densities  $\{\rho _{i}(\mathbf{r})\}$.
The free energy functional is commonly decomposed into the sum of three
contributions, namely into the ideal, %
the hard-sphere,
and the
electrostatic
terms. Of course, one can also
include here an additional term, due to possible association
of ions.
 Various formulations of the DFT have been
discussed in the literature.  As we  have already stressed, a
majority of the DFTs have followed perturbational second-order
expansion of the electrostatic free energy functional with respect
to a bulk homogeneous fluid~\cite{13,17,18,19,21}.

Recently, Gillespie et al.~\cite{24,25} proposed a version of the
electrostatic free energy functional that replaces a uniform
reference system with a suitably chosen position-dependent
reference fluid. The inhomogeneous reference fluid densities are
then computed from the local densities by a self-consistent
iteration procedure. Actually, Gillespie et al.~\cite{24,25}
proposed a reference fluid density functional which permits to
construct a reference model that
 locally satisfies electroneutrality condition and has
the same ionic strength at every point as the inhomogeneous fluid
in question. Such a construction of the reference fluid permits to
apply the expression for the
electrostatic contribution to the free energy, which results from the
 equation of state for a bulk ionic system
and makes it unnecessary to employ the direct correlation functions as input
quantities. This kind of approach was proposed by us to study
 liquid-vapor transitions in RPMs confined to slit-like pores~\cite{21d,pizio}.
The results of our approach  reasonably well reproduce the
structure of ionic fluids at a charged wall at sufficiently high
temperatures and correctly predict~\cite{reszko} the temperature
dependence of the double layer capacitance, in agreement with
experiments and simulation data~\cite{boda}.
 Quite recently, the theory was also extended to include the effects of association
 between unlike ions~\cite{pizio1}.
 However, the applications of the approach described above would not be complete
without exploring liquid-vapor interface of ionic fluid.
Therefore, in this communication we intend to apply the theory
of~\cite{pizio,reszko,pizio1} to the study of
 liquid-vapor interface of the RPM. We calculate the density profiles and the
 interfacial tension. Moreover, we would like to investigate the effect of
association leading to the formation of ion pairs and whether
this type of effects improves the description of the interfacial properties
similarly to the bulk structure and thermodynamics.

\section{Theory}

We consider a binary mixture of ionic species. The symbols
$d$, $Z_{i}=q_i/e$ and $\mu _{i}$ denote, respectively, the hard-sphere
diameter, valence of ions and the chemical potential of species $i=1,2$,
and $e$ is the electron charge.
The interaction between the ions is
\begin{equation}
u_{ij}(r)=\left\{
\begin{array}{ll}
\infty , & r<1, \\
{e}^{2}Z_{i}Z_{j}/\varepsilon r, & r>1,
\end{array}
\right.   \label{1}
\end{equation}
where $r$ is the distance between a pair of ions.
 We also assume that $Z_{1}=-Z_{2}=1$ and that the dielectric
constant $\varepsilon $ is uniform throughout the entire system.

The DFT we use in this work is identical to that described
in~\cite{pizio,pizio1} and therefore we present only its brief
description. If there is no external potential field the grand
potential  of the system  is written in the form,
\begin{equation}
\Omega =\int \rd\mathbf{r}f[\{\rho _{i}\}]-\sum_{i=1,2}\int \rho _{i}(z)\mu _{i}\rd\mathbf{r}.
\end{equation}
According to the usual density functional treatment, the free
energy density functional, $f[\{\rho _{i}\}],$ is decomposed into
ideal, hard sphere, electrostatic and associative terms $ f[\{\rho
_{i}\}]=f_{\mathrm{id}}[\{\rho _{i}\}]+f_{\mathrm{hs}}[\{\rho
_{i}\}]+f_{\mathrm{el}}[\{\rho _{i}\}] + f_{\mathrm{as}}[\{\rho
_{i}\}]$. The ideal term is $ f_{\mathrm{id}}[\{\rho
_{i}\}]=\sum_{i=1,2} [\rho _{i}(z)\ln \rho _{i}(z)-\rho _{i}(z)]$,
whereas for the hard-sphere contribution we apply an expression
resulting from a recent version~\cite{28} of the Fundamental
Measure Theory~\cite{15,16}, with the free energy density
consisting of terms dependent on scalar and on vector weighted
densities, $n_{\alpha }(\mathbf{r})$ ($\alpha =0,1,2,3$) and
$\mathbf{n}_{\alpha }(\mathbf{r})$ ($\alpha =V1,V2$)~\cite{28},
\begin{equation}
f_{\mathrm{(hs)}}=-n_{0}\ln (1-n_{3})+\frac{n_{1}n_{2}-\mathbf{n}_{V1}\cdot \mathbf{n}%
_{V2}}{1-n_{3}}+n_{2}^{3}(1-3\xi ^{2})\frac{n_{3}+(1-n_{3})^{2}\ln (1-n_{3})%
}{36\pi n_{3}^{2}(1-n_{3})^{2}}\;,  \label{eq:7}
\end{equation}
where $\xi
(\mathbf{r})=|\mathbf{n}_{V2}(\mathbf{r})|/n_{2}(\mathbf{r})$. The
definitions of  weighted densities, $n_{\alpha }(\mathbf{r}),$
$\alpha =0,1,2,3,V1,V2$, are given in~\cite{15,16}.

The electrostatic contribution is evaluated from the approach
described in~\cite{pizio,pizio1}, according to which
\begin{equation}
f_{\mathrm{el}}[\{\rho _{i}\}]/kT=-\frac{e^{2}}{T^{\ast }}[Z_{1}^{2}\bar\rho_1
+Z_{2}^{2}\bar\rho _{2}]\frac{\Gamma(\{\bar\rho_i\}) }{1+\Gamma(\{\bar\rho_i\}) d }
+\frac{\Gamma ^{3}(\{\bar\rho_i\})}{3\pi }\,,
\end{equation}
where $\bar{\rho}_{i}(z)$ denote suitably defined inhomogeneous
average densities of a reference fluid. The form for
$f_{\mathrm{el}}[\{\bar{\rho}_{i}(z)\}]$ results from the MSA
equation of state evaluated via the energy
route~\cite{1,2,3,7,WL,WL1}. In the above $ \Gamma(\{\bar\rho_i\})
=(\sqrt{1+2\kappa(\{\bar\rho_i\}) d }-1)/2d, $ where
$\kappa(\{\bar\rho_i\}) $ denotes the inverse Debye screening
length, $ \kappa ^{2}(\{\bar\rho_i\})=(4\pi e^{2}/T^{\ast
})[Z_{1}^{2}\bar\rho _{1}+Z_{2}^{2}\bar\rho _{2}]. $

The reference fluid densities $\{\bar\rho_{i}\}$ are evaluated by
employing the approach of Gillespie et al.~\cite{24,25}. In the
case of gas-liquid interface, the spatial symmetry of the RPM
implies that the density profiles of the two ionic species should
be the same. Therefore, in the presently considered situation,
 the approach of Gillespie
et al.~\cite{24,25} is simplified  and the reference fluid
densities are given by
\begin{equation}
\bar{\rho}_{i}(z)=\int \rho _{i}(z)W_{i}(|\mathbf{r}-\mathbf{r^{\prime }}%
|)\rd\mathbf{r^{\prime }},
\end{equation}
where the weight function $W_{i}(r)$
is just a normalized step function
\begin{equation}
W_{i}(|\mathbf{r}-\mathbf{r^{\prime }}|)=\frac{\theta (|\mathbf{r}-\mathbf{%
r^{\prime }}|-R_{f}(\mathbf{r^{\prime }}))}{(4\pi /3)R_{f}^{3}(\mathbf{%
r^{\prime }})}\,.
\end{equation}
The radius of the sphere over which the averaging is performed, $R_f$,
is approximated by  the
 ``capacitance'' radius, i.e., by the ion
radius plus the screening length
\begin{equation}
R_{f}(\mathbf{r})=\frac{d}{2}+\frac{1}{2\Gamma (\{\bar{\rho}_{i}(\mathbf{r}%
)\})}\,.
\end{equation}
The evaluation of $R_{f}$
requires an iteration procedure. This iteration loop should be carried out
in addition to the main iteration procedure used to evaluate
 the density
profiles, see for details~\cite{24,25}.

 Finally, the associative contribution, $f_{\mathrm{as}}$,
is  formulated at the level of the first-order thermodynamic
perturbation theory~\cite{pizio1,we}
\begin{equation}
f_{\mathrm{as}}/kT=\sum_{i=1,2} \bar{\rho}_{i}\left[\ln \alpha (\{\bar{\rho}_{i}\})+\frac{1%
}{2}-\frac{\alpha (\{\bar{\rho}_{i}\})}{2}\right],
\end{equation}
where $\alpha $ is the degree of dissociation according to the mass
action law,
$
2{\alpha}={1-K{(\bar{\rho}_{1}+\bar{\rho}_{2})\alpha ^{2}}}.
$
The association constant $K\equiv K(\{\bar{\rho%
}_{i}\},T)$ is a product of solely temperature dependent term,
$K^{0}$  and $K^{\gamma } $, $K=K^{0}K^{\gamma }$. The constant
$K^{0}$, is~\cite{13a}
\begin{equation}
K^{0}=8\pi d^{3}\sum_{m=2}^{\infty} \frac{(T^{\ast })^{-2m}}{(2m)!(2m-3)}\,,
\end{equation}
whereas $K^{\gamma }$ follows from the so-called simple
interpolation scheme~\cite{7,pizio1,f2}
\begin{equation}
K^{\gamma }=\frac{(1-\bar{\eta}/2)}{(1-\bar{\eta})^{3}}\exp \left[ -\frac{%
\Gamma d}{T^{\ast }}\frac{(2+\Gamma d)}{(1+\Gamma d)^{2}}\right] .
\end{equation}

The theory presented above uses different weighted densities to evaluate
the hard sphere and electrostatic free energy contributions.
Note that the idea of
using different weighted densities to evaluate these contributions
 has also been employed
by Patra et al.~\cite{18,19}.

The density profile  is obtained by minimizing the excess grand
potential functional $\Delta \Omega =\Omega -\Omega _{b}$,
\begin{equation}
\frac{\delta \Delta \Omega }{\delta \rho _{i}(z)}=0,{\ \ \mathrm{for=1,2}},
\end{equation}
where
$\Omega_b$ is the grand potential of a bulk uniform system of density $\{\rho_{b,i}\}$.

The evaluation of the density profiles requires the knowledge
of the densities of both coexisting phases. Therefore, we have precisely evaluated
the bulk phase diagram prior to the density profile calculations.
Next, the local density calculations were carried out assuming that for $z\leqslant -L_z$ and
for $\geqslant L_z$ the density of the fluid is equal to the bulk densities of
gaseous and liquid phases, respectively. The value of $L_z$ was evaluated at
each temperature by testing the convergence of the density profiles towards the
final solutions.
We have started with $L_z=30d$ and the consecutive runs were carried out doubling
the value of $L_z$ for each new run. All the integrations were performed
using Simpson method with the grid size of 0.01d. The convergence
criterion definitly states that the maximum difference between two consecutive iterations
must be smaller than 1E-8 percent.
  The knowledge of the density profiles of ions permits to calculate the
excess grand thermodynamics potential per unit
surface area, $A$, i.e., the surface tension
$\gamma=\Delta\Omega/A$.

\section{Results and discussion}

Figure~1 shows the phase diagram in the density-temperature plane.
The phase diagram resulting from the MSA energy equation of state
has been presented in several previous works, cf.~\cite{7,Groh}.
Nevertheless,
 we have decided to include it here for completeness of the study.
The solid line corresponds to the system without association (i.e. the associative
free energy term is neglected in the grand thermodynamics potential functional),
whereas the dashed line describes phase behavior of the model with the association
effects included.
Inclusion of  chemical association into the theory
provides a mechanism for the formation of ionic pairs. The fraction of pairs
depends on density, on temperature and on the association constant.
From physical point of view, the formation of pairs of oppositely charged ions,
 alters the effective interactions between all the particles.
In effect,  the critical
temperature of liquid-vapor transition, $T^*_{\mathrm{c}}\approx 0.0745$,  becomes lower compared to
the critical temperature of the model without association, $T^*_{\mathrm{c}}\approx 0.0786$.
The liquid-vapor phase diagrams are strongly asymmetric.
At low temperatures the vapor phase is very dilute. For example, for the system
 without association at $T^*=0.05$ the vapor density is
 $\rho^*_b\approx 1.094E-5$, whereas the density of the coexisting liquid
 phase is $\rho^*_b\approx 0.2135$.
\begin{figure}[ht]
\begin{center}
\includegraphics[width=0.45\textwidth]{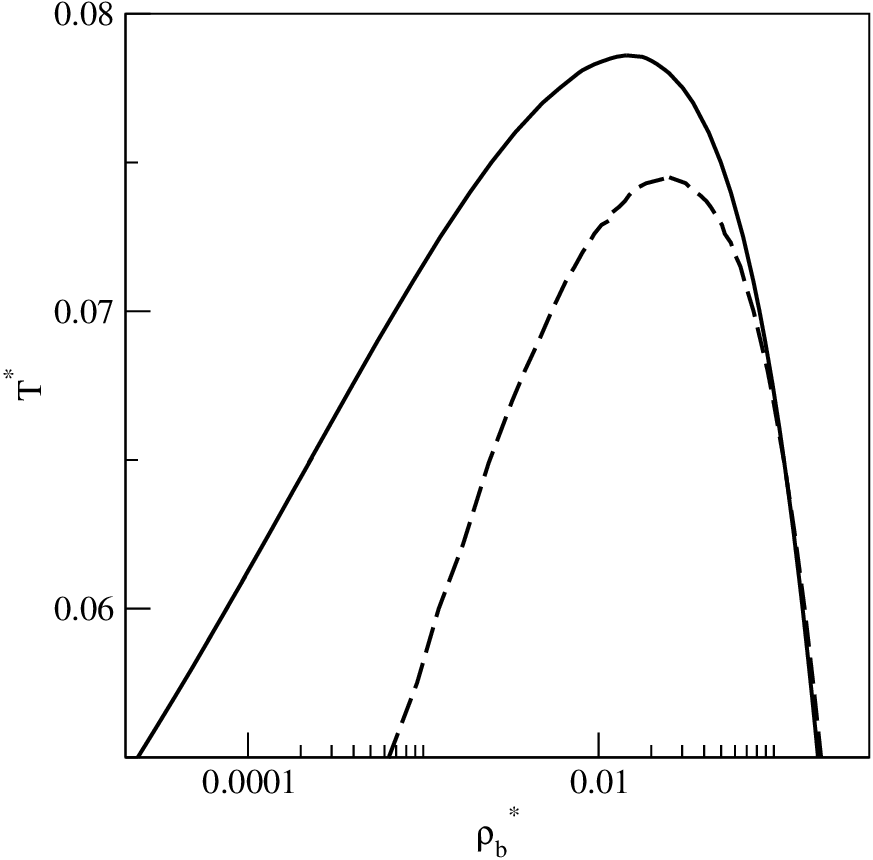}
\caption{Phase diagram resulting from MSA approach and the energy equation
of state. Solid and dashed lines denote the results for the
model without and with association.}
\end{center}
%\label{fig:1}
%}
\end{figure}
\begin{figure}[!h]
\begin{center}
\includegraphics[width=0.45\textwidth]{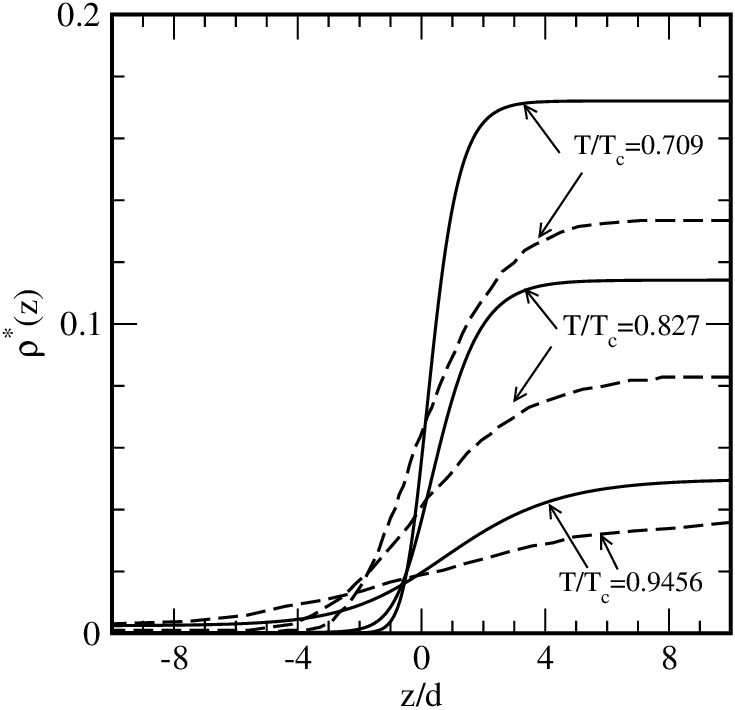}
\caption{A comparison of the total density profiles,
 $\rho^*(z)=(\rho_1(z)+\rho_1(z))d^3$ resulting from the DFT without association
 theory (solid lines) and from MSA1 theory of Groh et al.~\cite{Groh}
 (dashed lines). The temperatures
$T/T_{\mathrm{c}}$ are given in the figure.}
\end{center}
\end{figure}

 First, we consider the system without association.
 The evaluation of the liquid-vapor
  density profiles was carried out assuming that the limiting values $\rho_i(-L_z)$ and $\rho_i(L_z)$
  are equal to the MSA densities of coexisting phases. Of course,
  the agreement of our results
   with the results published previously~\cite{Groh,23}
 essentially depends on the agreement between the bulk
  phase diagrams.  One should keep
  in mind the above remark while analyzing figure~2, where we show a comparison
  of our results with those of Groh et al.\cite{Groh}. The latter
  profiles were obtained from
  a different bulk theory (called by them ``the MSA1 approach''~\cite{Groh}), which yields
  the value of the critical temperature, $T_{\mathrm{c}}^*\approx 0.0846$,
    different from that obtained by MSA theory. Note that a comparison of the
  entire phase diagrams resulting
  from the MSA and MSA1 approaches in given in figure~1 of~\cite{Groh}.

\begin{figure}[ht]
\begin{center}
\includegraphics[width=0.45\textwidth]{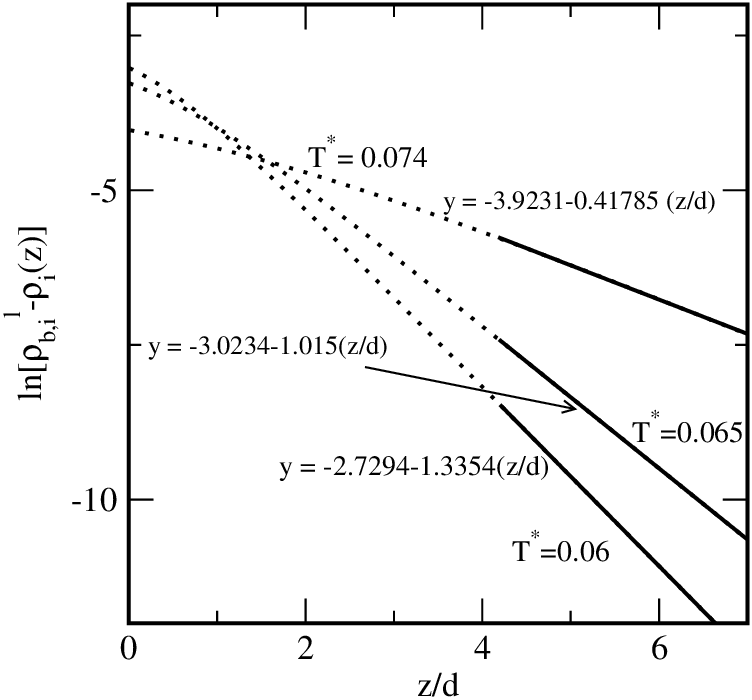}
\caption{The convergence of the liquid parts of the single species
density profiles towards the bulk liquid equilibrium density at three
different temperatures given in the figure. Figure  also contains
the equations for straight lines (marked by solid lines) approximating the
data,
computed from density functional theory. Here $y=\ln[\rho){b,i}^{\mathrm{l}}-\rho_i(z)]$
and $ \rho_{b,i}^{\mathrm{l}}$ is the single species
density of an ionic liquid at  coexistence. The calculations are for
the model without association.}
\end{center}
\end{figure}

  The way in which the density profile approaches the asymptotic bulk
  values has been discussed by Groh et al.~\cite{Groh}. We have
   checked that
  the decay of the function
   $\rho_{b,i}^{\mathrm{l}}-\rho_i(z)$ (where $\rho_{b,i}^{\mathrm{l}}$ is the
  liquid-phase bulk density) with $z\to L_z$ is exponential, i.e.
  $\ln[\rho_i(z)-\rho_{b,i}^{\mathrm{l}}]\propto z/\xi$.
  This point is illustrated in figure~3, where we have plotted
   $\ln[\rho_{b,i}^{l}-\rho_i(z)]$ versus $z/d$. At larger distances
  this dependence is fairly well approximated by a straight line with the
  correlation coefficient, $R^2$, very close to unity.

\begin{figure}[!h]
\begin{center}
\includegraphics[width=0.43\textwidth] {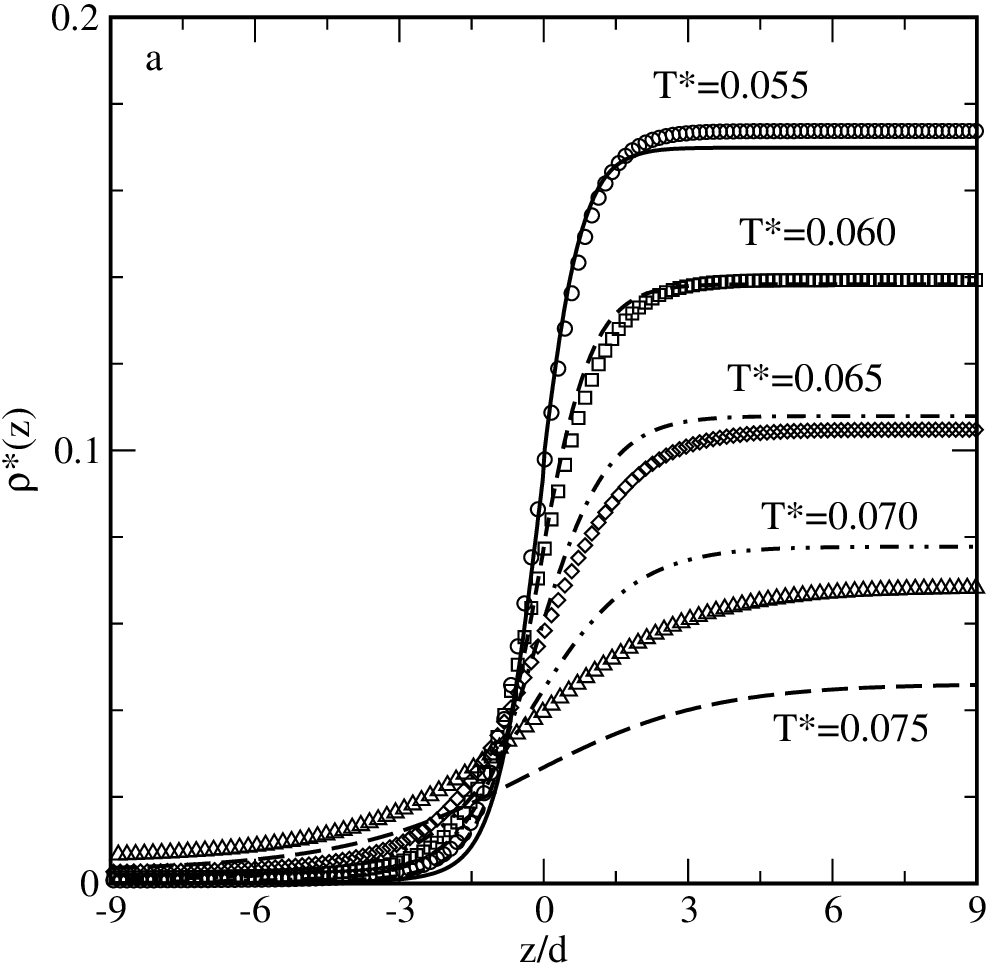}
\includegraphics[width=0.43\textwidth] {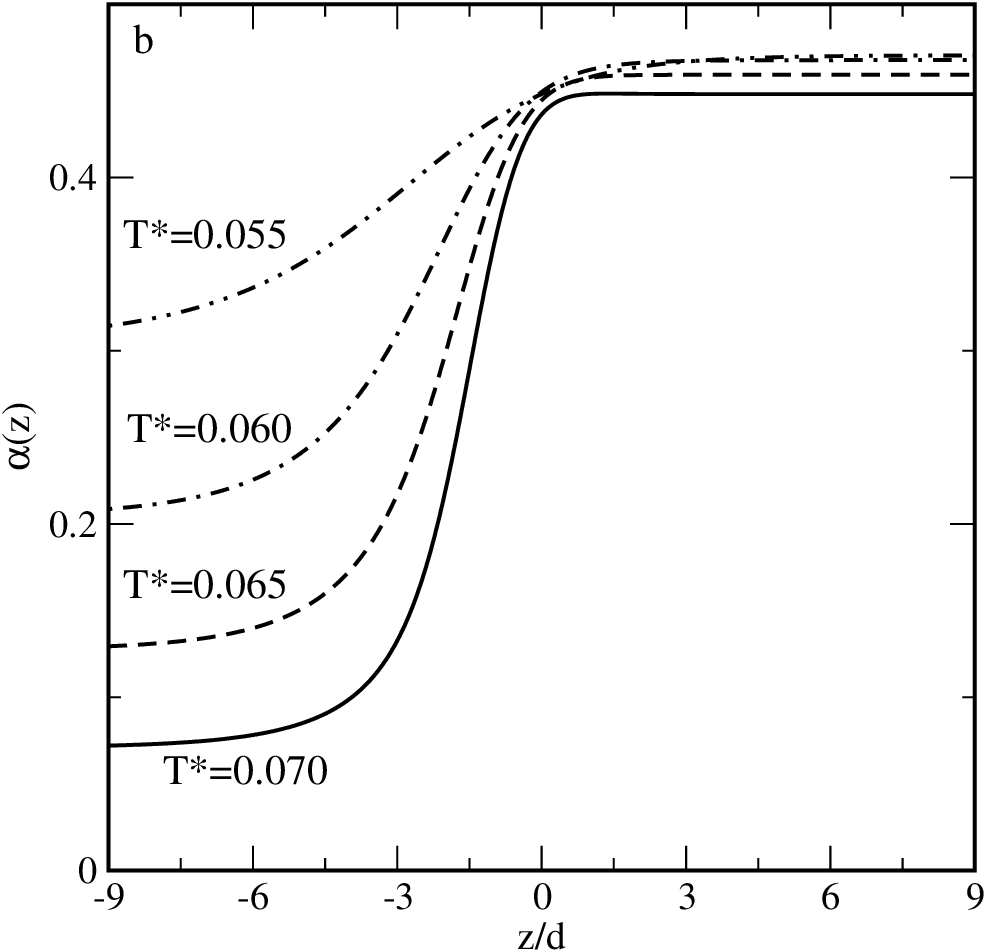}
\caption{Part {\it a}. A comparison of the total density profiles for the models without (solid lines) and
with association (symbols). Part {\it b}. Dissociation degrees across the vapor-liquid interface.
The temperatures are given in the figure.
}
\end{center}
\end{figure}
\begin{figure}[!h]
\begin{center}
\includegraphics[width=0.45\textwidth] {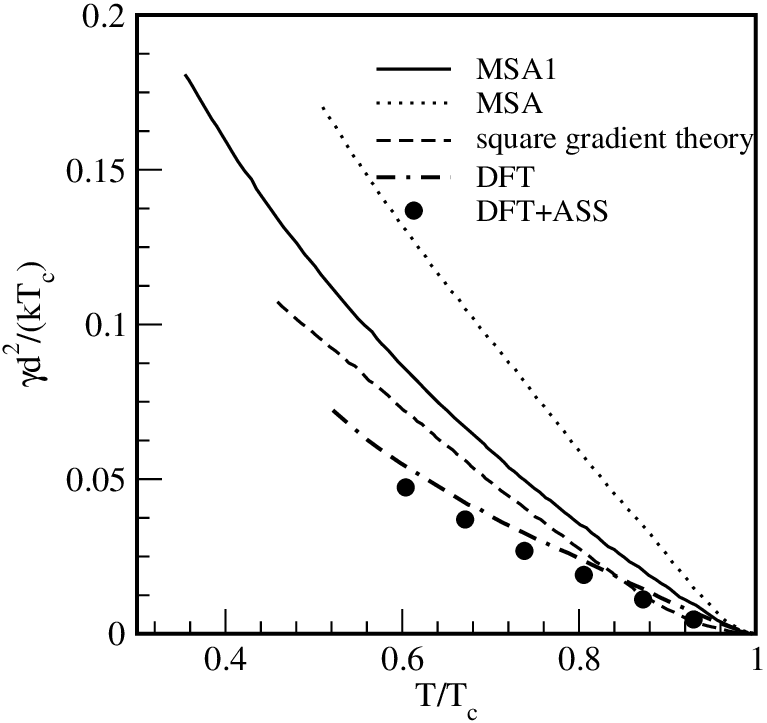}
\caption{The dependence of the surface tension on the temperature.
Curves labelled MSA, MSA1 were evaluated by Groh et
al.~\cite{Groh}, the curve labelled ``square gradient theory'' was
evaluated by Telo da Gama et al.~\cite{Gama}, the curve
abbreviated as DFT was obtained for the model without association,
whereas black circles -- for the model with the association
effects included.}
\end{center}
\end{figure}
  We now consider the effect of the association, cf. figure~4. Part {\it a}
shows a comparison of
  the total density profiles for the systems with (symbols) and without
  association (lines) at different temperatures. At low temperatures the
  liquid density for the system with the association included is higher
  than for the system with the association effects switched off, cf. figure~1.
  The profiles for the model with association are more ``smeared out'', but
  this is the effect of lowering the critical temperature due to association effects.
  Similarly to the case of a system without association  we have
  found exponential decay of the function $\rho_{b,i}^{\mathrm{l}}-\rho_i(z)$  with $z\to L_z$
  (the corresponding plot was omitted for the sake of brevity).
  Figure 4b shows the profiles of $\alpha(z)$ (the phase diagram in the
  $\alpha$-$T$ plane has been already presented in~\cite{pizio1})
  We stress that the interfacial  region  of $\alpha(z)$ is shifted towards rarefied
  phase compared to the density profiles. It means that the position
of the ``pseudo-Gibbs'' dividing surface,
  which can be introduced for the dissociation degree in
  analogy to the usual liquid-vapor dividing surface is also shifted
  compared to the position of the usual dividing surface (which is located at $z=0$).
  Note that similar shift of the dissociation degree was also observed
  in the case of liquid-vapor interfaces of non-ionic fluids~\cite{bor}.

\looseness=-1  The surface tension obtained from the density functional theory
  and from the MSA1 and MSA approaches, developed by Groh et al.~\cite{Groh},
  as well as from the square gradient theory~\cite{Gama} are shown
  in figure~5. We have plotted here
  the ratio $\gamma d^2/kT_{\mathrm{c}}$ versus the temperature
  reduced by the critical temperature, $T/T_{\mathrm{c}}$. Except for the
  temperatures very close to the critical temperature, the present
  theory predicts lower values of the surface tension than all
  the rest approaches. The associative free energy terms still lower
  the value of $\gamma d^2/kT_{\mathrm{c}}$. Note that the MSA theory grossly underestimates
  the densities of the coexisting liquid phase. This can suggest
  that the values of the surface tension are also underestimated.
  On the other hand, the value of the critical temperature is
  overestimated by the MSA approach, and hence can lead to too high
  values of the interfacial tension.

\begin{figure}[ht]
\begin{center}
\includegraphics[width=0.45\textwidth] {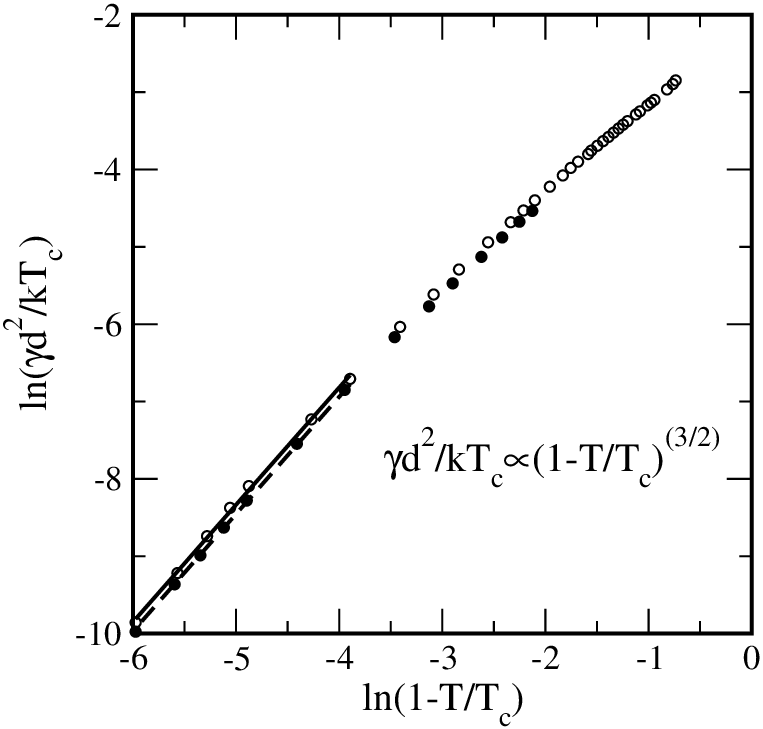}
\caption{The decay of the surface tension on approaching the
critical temperature. Solid line and empty circles are for
the model without association, whereas dashed line and  filled circles -- for
the model with association. The straight lines slope is 3/2.}
\end{center}
\end{figure}
  Near critical temperature, both models (i.e., the model with and
  without association) yield the standard
  mean-field critical behavior, i.e., $\gamma d^2/kT_{\mathrm{c}}\propto
  (1 -T/T_{\mathrm{c}})^{(3/2)}$ (see figure~6). As the temperature
  approaches the critical point,
  the theory yields a nearly linear variation of $\ln[\gamma d^2/kT_{\mathrm{c}}]$
  with $\ln[(1 -T/T_{\mathrm{c}})]$ and the slope of the  straight line
  approximating the obtained data is
  almost perfectly equal to (3/2).

\begin{figure}[!h]
\begin{center}
\includegraphics[width=0.45\textwidth] {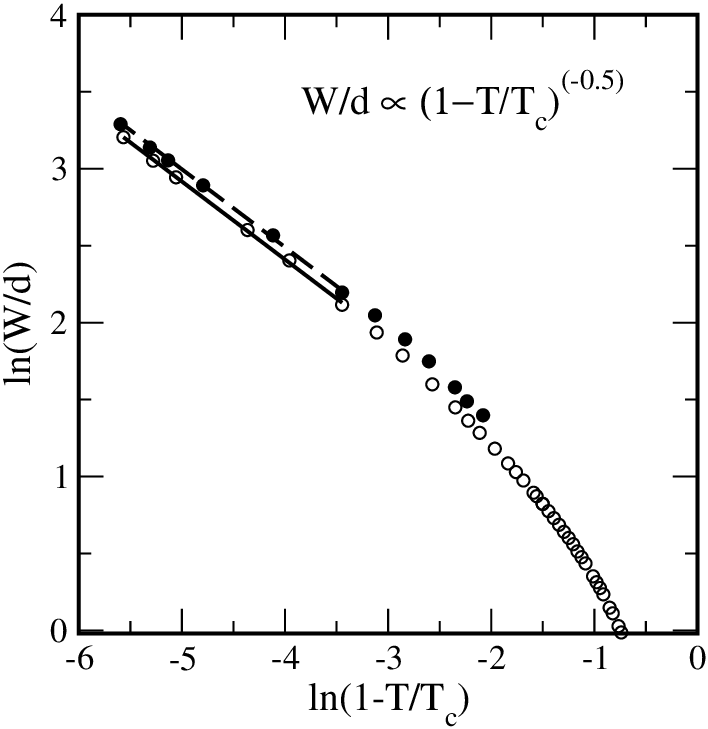}
\caption{The divergence  of the interfacial width $W$,
on approaching the
critical temperature. The straight lines slope is $-1/2$.
Abbreviations are the same as in figure~6.}
\end{center}
\end{figure}
  The width of the interfacial zone can be characterized by
the parameter $W$, defined as~\cite{Fischer}
\begin{equation}
W=-[\rho(z=L_z)-\rho(z=-L_z)]
\left[{\frac{\rd\rho(z)}{\rd z}}\right]^{-1}_{z=z_0},
\end{equation}
where $z_0$ is given by $\rho(z_0)=(1/2)[\rho_s(z=L_z)+\rho(z=-L_z)]$
and $\rho(z=L_z)$ and $\rho(z=-L_z)$ are the total densities
of the coexisting liquid and vapor phases.
  Obviously, the interface
 becomes wider as the temperature increases. The effects of
 association slightly increases the interface width.
 As the temperature approaches the critical
 temperature, the interfacial width diverges. To investigate
 the character of this divergence, we have plotted the
 values of $\ln W/d$  versus $\ln(1-T/T_{\mathrm{c}})$, see
 figure~7.
For $T$ approaching the critical temperature,
 the dependence of $\ln W/d$  versus $\ln(1-T/T_{\mathrm{c}})$ is linear.
 We have found the following scaling:
 $W^*\propto (1-T/T_{\mathrm{c}}(M))^{-0.5}$. The obtained value of the
 exponent (0.5) is characteristic of mean-field type theories
~\cite{Groh}. The scaling of $W$ is consistent with the scaling of
the surface tension.

\begin{figure}[ht]
\begin{center}
\includegraphics[width=0.45\textwidth] {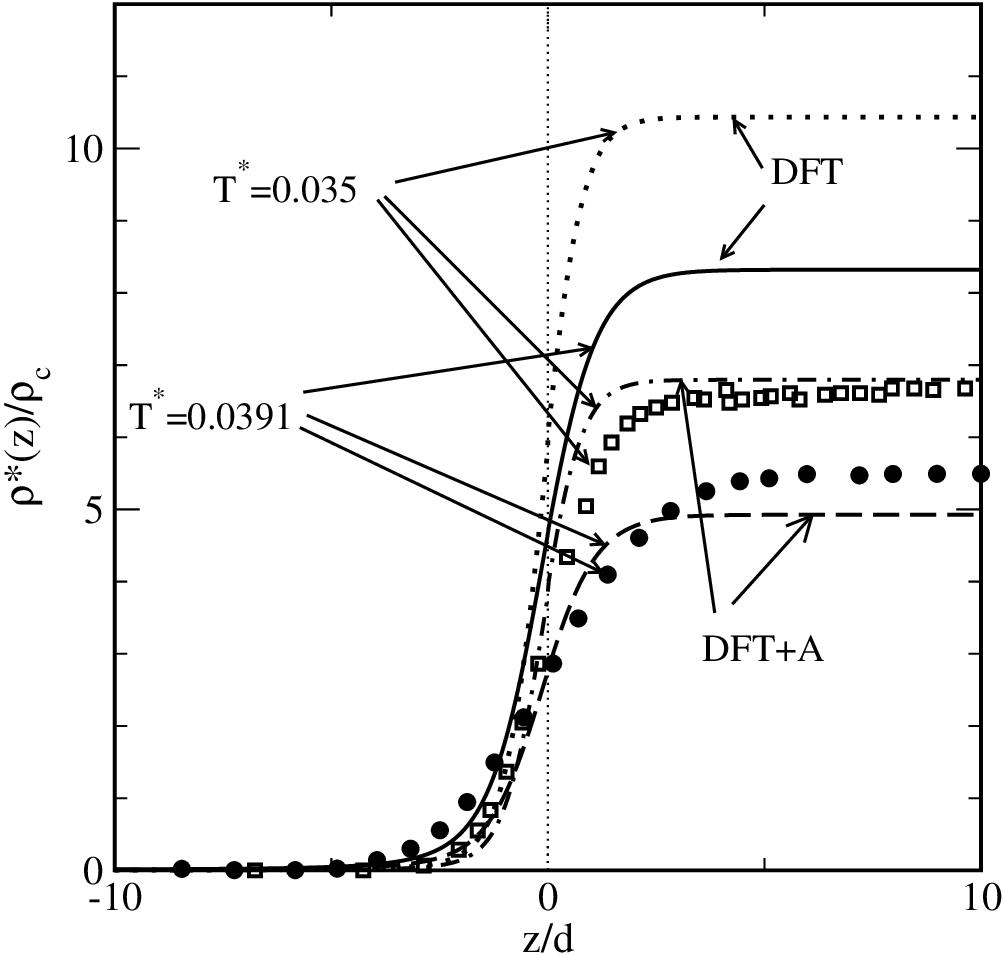}
\caption{A comparison of the total density profiles for the models without
association (solid and dotted lines) and
with association (dashed and dash-dotted lines) with computer
simulation data. The simulation
results for the profiles (see details in the text) are shown by symbols
\cite{alejandre1,alejandre2}. The temperatures are given in the figure.}
\end{center}
\end{figure}
Our discussion involved the results of different, but solely
theoretical approaches. However, it must be clarified how
accurate our approaches are (with and without association)
compared to the available computer
simulation data. In particular, Alejandre and co-workers
investigated the liquid-vapor interface for the RPM
 using molecular dynamics simulation method supplemented by
certain specific technical peculiarities
 dealing specifically with the RPM~\cite{alejandre1,alejandre2}.
These authors presented the density profiles, $\rho^{*}(z)$ at
four reduced temperatures, $T^*$=0.035, 0.038, 0.040 and
0.043~\cite{alejandre1}. On the other hand, a single density
profile at $T^*=0.391$ relevant to our study was given in
\cite{alejandre2} (this profile is for the soft primitive model
but with parameters permitting comparison with the RPM). The
authors have not performed estimates of the critical parameters
coming from their method of study of the RPM. Thus, they somewhat
arbitrarily used the critical temperature obtained in~\cite{6},
$T^*_{\mathrm{c}}=0.0492$, for their purposes. We have rescaled
the temperature and density with respect to the critical
parameters of both theoretical approaches (with and without
association), respectively, and present here a comparison of the
density profiles with computer simulation data in figure~8. Note
that the critical density given in~\cite{6} was used to rescale
the simulation data. We observe that the \textit{rescaled} liquid
phase density at coexistence coming from the DFT for the model
with ion pairing is in reasonable agreement with the simulation
data. The DFT for the model without association greatly
overestimates these liquid densities. However, the width of the
interfacial region is not described well. Theoretical approaches
yield narrower interface width for two reduced temperatures in
question. It is difficult to attribute this specific inaccuracy to
the particular term of the free energy functional. However, it
seems that the mean field consideration requires improvement in
order to better describe interparticle correlations  and hopefully
to obtain a more adequate interface width.

As concerns the relationship between the simulation data for the
surface tension and the results of theoretical approaches we would
like to comment on the following points. A set of computer
simulation results for the surface tension were put together with
the predictions of some theoretical approaches~\cite{Gama,Groh,23}
(that were discussed in the introductory part of our
communication) on arbitrary temperature scale, see figure~2
of~\cite{alejandre1}. Apparently, the theories yield the surface
tension of the same magnitude as computer simulation data. This
permitted to conclude that the surface tension from this set of
theories~\cite{Gama,Groh} is overestimated because the effect of
ion association is not taken into account and solely the theory
that accounts for association~\cite{23} is in good agreement with
simulations. Then, it would seem that our approach
 taking  ion pairing into account is the best, see figure~5.
Moreover, a comparison of the theories, simulations and
experimental data was performed~\cite{alejandre2} using an
arbitrary scale. Such an idyllic picture is, however, entirely
destroyed, if one rescales the reduced temperature scale by the
critical temperature of each theory and of simulations. This is
actually an adequate comparison. According to~\cite{alejandre1}
the simulations
 yield, e.g. $\gamma d^2/kT_{\mathrm{c}}$=0.3 at $T/T_{\mathrm{c}}$=0.6. Such a value
is very much higher than the results of any theory, c.f. figure~5,
at this particular temperature. Moreover, this means that the
inclusion of association effects makes things worse than from any
other theory. This is difficult to accept, however. According to
the more recent simulation work of the same
authors~\cite{alejandre2} it is difficult to draw a definite
conclusion about the dependence of the surface tension on
temperature for the model in question due to the statistical error
associated with the value for surface tension. Moreover, it has
been discussed~\cite{alejandre2} that the inflection point at
$T^*=0.04$ (again it is a reduced temperature of simulation) can
be present on the dependence of surface tension on temperature as
a result of chemical association of ions in the RPM in close
similarity to hydrogen bonded fluids. In our opinion, additional
work, both theoretical and simulational, is necessary to obtain
definite answers about the behavior of surface tension on
temperature for the model in question and the related models. From
theoretical viewpoint one needs a theory that yields a more
adequate shape of the bulk coexistence envelope. Then, the free
energy expression coming from this approach, if available, would
contribute to the development of better density functional
approaches for inhomogeneous ionic fluids.

To summarize briefly, in this work we have applied  density
functional theory to the study of the liquid-vapor interface of a
RPM fluid. Of course, the theory we use here is \textit{ad hoc}.
However, our previous calculations~\cite{pizio,reszko} have shown
that it reproduces reasonably well the structure of a fluid at a
charged and uncharged wall, predicts the dependence of the
capacitance of the double layer on the temperature and  yields
``capillary evaporation'' phase diagrams for confined ionic
systems~\cite{pizio,reszko,pizio1}. We have shown that the
inclusion of ion pairing effect leads to a better agreement of the
density profiles of ions across the liquid-vapor interface. A more
sophisticated approaches~\cite{7}, compared to the present one,
including the effects of ionic association can lead to even better
agreement of the coexistence envelope with simulations and of the
density profiles across the liquid-vapor interface. We plan to
extend our study in this respect in the nearest future. However,
the inaccuracies of the values for the liquid phase density at
coexistence at different temeperatures prevent us from obtaining a
reasonable agreement with the available simulation data for the
surface tension. It seems, however, that a more ample set of
simulation data using different techniques is necessary.
Nevertheless, the theories of this study yield a decreasing
surface tension with increasing temperature as one intuitively
would expect. Still we are not able to conclude that the
theoretical approach is entirely successful both for the
fluid-solid and fluid-fluid interfaces and look forward to
improving it in future work.

\ukrainianpart

\title{Міжфазна границя рідина-пара обмеженої примітивної моделі іонних плинів з теорії функціоналу густини}

\author{А. Патрикеєв\refaddr{label1},
С. Соколовскі\refaddr{label1},
О. Пізіо\refaddr{label2}}

\addresses{
\addr{label1}Університет ім. Марії Складовської-Кюрі, Люблін, Республіка Польща,
\addr{label2}Інститут хімії УГАМ, Койокан, Мексика}

\date{Отримано 13 вересня 2010р., в остаточному вигляді~--- 23 листопада 2010р.}

\makeukrtitle

\begin{abstract}
\tolerance=3000%
Ми досліджуємо міжфазну границю рідина-пара обмеженої примітивної моделі іонного плину, використовуючи теорію функціоналу густини. Застосована теорія включає електростатичний вклад у функціонал вільної енергії, який виникає з енергетичного рівняння стану для об'єму, вклад від середньо-сферичного наближення  для обмеженої примітивної моделі, а також асоціативний вклад, який виникає в результаті  врахування утворення іонних пар. Ми порівнюємо профілі густини і значення поверхневого натягу з результатами попередніх теоретичних підходів.

\keywords функціонал густини, адсорбція, ланцюжки, кристали
\end{abstract}


\begin{thebibliography}{99}

\bibitem{DH} Debye P.W., H\"uckel E., Phys. Z., 1923, {\bf 24}, 185.

\bibitem{NNN}  Moreira A.G., Netz R.R., European Phys. J. D, 2001, {\bf 13}, 61; \bibdoi{10.1007/s100530170287}.


\bibitem{1}  Blum L., {J. Chem. Phys.}, 1974, \textbf{61}, 2129; \bibdoi{10.1063/1.1682224}.

\bibitem{2}  Blum L.,  {Molec. Phys.}, 1975, \textbf{30}, 1529; \bibdoi{10.1080/00268977500103051}.

\bibitem{2a}  S\'anchez-Diaz L.E., Vizcarra-Rend\'on A.,
Medina-Noyola M., J. Chem. Phys., 2010, \textbf{132}, 234506;\\
\bibdoi{10.1063/1.3455336}.

\bibitem{WL} Waisman E., Lebowitz J.L., {J. Chem. Phys.}, 1970, {\bf 52}, 4307; \bibdoi{10.1063/1.1673642}.

\bibitem{WL1} Waisman E., Lebowitz J.L., {J. Chem. Phys.}, 1972, {\bf 56}, 3093; \bibdoi{10.1063/1.1677645}.




\bibitem{3} Panagiotopoulos A.Z., {Fluid Phase Equilibria}, 1992, \textbf{76}, 97; \bibdoi{10.1016/0378-3812(92)85080-R}.

\bibitem{4}  Caillol J.N., {J. Chem. Phys.}, 1994, \textbf{100}, 2161; \bibdoi{10.1063/1.466513}.

\bibitem{5}  Romero-Enrique J.M., Orkoulas G.,
 Panagiotopoulos A.Z., Fisher M.E., {Phys. Rev. Lett.}, 2000, \textbf{85}, 4558; \bibdoi{10.1103/PhysRevLett.85.4558}.

\bibitem{6} Yan Q.L., de Pablo J.J.,  {J. Chem. Phys.}, 2001, \textbf{114}, 1727; \bibdoi{10.1063/1.1335653}.

\bibitem{SYM} Orkoulas G., Panagiotopouloas A.Z.,
{J. Chem. Phys.}, 1994, {\bf 101}, 1452; \bibdoi{10.1063/1.467770}.

\bibitem{7}  Jiang J., Blum L., Bernard O.,
 Prausnitz J.M.,  Sandler S.I.,
{J. Chem. Phys.}, 2002, \textbf{116}, 7977; \\
\bibdoi{10.1063/1.1468638}.

\bibitem{new7} Valeriani C., Camp P.J., Zwanikken J.W., van Roij R., Dijkstra M.,
J. Phys. Condens. Mat., 2010,  \textbf{22}, 104122; \bibdoi{10.1088/0953-8984/22/10/104122}.

\bibitem{new8}
Diehl A., Panagiotopoulos, A.Z.,  J. Chem. Phys., 2006, \textbf{124}, 194509; \bibdoi{10.1063/1.2192498}.

\bibitem{8}  Stell G., {J. Phys.: Condens. Mat.}, 1996, \textbf{8}, 9329; \bibdoi{10.1088/0953-8984/8/47/024}.

\bibitem{9}  Fisher M., Levin Y.,  {Phys. Rev. Lett.}, 1993, \textbf{71}, 3826; \bibdoi{10.1103/PhysRevLett.71.3826}.

\bibitem{hol1} Holovko M.F., {J. Mol. Liq.}, 2002, \textbf{96--97}, 65; \bibdoi{10.1016/S0167-7322(01)00327-0}.

\bibitem{LJ} Rowlinson J.S., Widom B., {Molecular Theory of Capillarity},
Claredon, Oxford, 1982.

\bibitem{Gama} Telo da Gama M.M., Evans R., Sluckin T.J., {Molec. Phys.}, 1980,
{\bf 41}, 1355; \\ \bibdoi{10.1080/00268978000103591}.

\bibitem{Gradient} Evans R., Sluckin T.J., {Molec. Phys.}, 1980,
{\bf 40}, 413; \bibdoi{10.1080/00268978000101581}.

\bibitem{Groh} Groh B., Evans R., Dietrich S.,
{Phys. Rev. E}, 1998, \textbf{57}, 6944; \bibdoi{doi:10.1103/PhysRevE.57.6944}.




\bibitem{13}  Mier y Teran L., Suh S.H., White H.S., Davis H.T.,
{J. Chem. Phys.}, 1990, \textbf{92}, 5087;
\\ \bibdoi{10.1063/1.458542}.

%

\bibitem{17} Kierlik E., Rosinberg M.L.,  {Phys. Rev. A}, 1991,
 \textbf{44}, 5025; \bibdoi{10.1103/PhysRevA.44.5025}.

\bibitem{18}  Patra C.N., Ghosh S.K.,
{Phys. Rev. E}, 1993, \textbf{47}, 4088; \bibdoi{10.1103/PhysRevE.47.4088}.

\bibitem{19}  Patra C.N., {J. Chem. Phys.}, 1999, \textbf{111}, 9832; \bibdoi{10.1063/1.480319 }.


\bibitem{21}  Boda D., Henderson D., Mier y Teran L., Soko\l owski S.,
{J. Phys. Condens. Mat.,} 2002, \textbf{14}, 11945;
\\ \bibdoi{10.1088/0953-8984/14/46/305}.

\bibitem{21a} Pizio O., Bucior K., Patrykiejew A., Soko\l owski S.,
{J.  Chem. Phys.}, 2005, \textbf{123}, 214902;
\\ \bibdoi{10.1063/1.2128701}.


\bibitem{21b} Goel T., Patra C.N., Ghosh S.K., Mukherjee T., Molec. Phys.,
2009, \textbf{107}, 19; \\ \bibdoi{10.1080/00268970802680497}.


\bibitem{21c} Goel T., Patra C.N., Ghosh S.K., Mukherjee T., J. Chem. Phys.,
2008, \textbf{129}, 154906; \\ \bibdoi{10.1063/1.2992525}.

\bibitem{21d}  Pizio O., Patrykiejew A., Sokolowski S., Condens. Matter Phys., 2004,
\textbf{7}, 779.

\bibitem{23} Weiss C.V., Schr\"oer W.,
{J. Phys. Condens. Mat.,} 2000, \textbf{12}, 2637; \bibdoi{10.1088/0953-8984/12/12/306}.


\bibitem{24}  Gillespie D., Nonner W.,  Eisenberg R.S.,
J. Phys. Condens. Mat., 2002, \textbf{14}, 12129;
\\ \bibdoi{10.1088/0953-8984/14/46/317}.

\bibitem{25}  Gillespie D., Nonner W., Eisenberg R.S.,
Phys. Rev. E, 2003, \textbf{68}, 031503;
\\ \bibdoi{10.1103/PhysRevE.68.031503}.


 \bibitem{pizio} Pizio O., Patrykiejew A., Soko\l owski S., J. Chem. Phys.,
2004, {\bf 121}, 11957; \bibdoi{10.1063/1.1818677}.


\bibitem{reszko} Reszko-Zygmunt J., Soko\l owski S., Henderson D., Boda D.,
J. Chem Phys., 2005, {\bf 122},  084504; \\
\bibdoi{10.1063/1.1850453}.

\bibitem{boda} Holovko M., Kapko V., Henderson D., Boda D.,
Chem. Phys. Lett., 2001, \textbf{341}, 363;
\\ \bibdoi{10.1016/S0009-2614(01)00505-X}.

\bibitem{pizio1} Pizio O., Soko\l owski S., J. Chem. Phys., 2005, {\bf 122}, 144707; \bibdoi{10.1063/1.1883165}.

\bibitem{28}  Yu X.Y., Wu J.Z., {J. Chem. Phys.}, 2002, \textbf{117}, 10156; \bibdoi{10.1063/1.1520530}.

\bibitem{15}  Rosenfeld Ya., {Phys. Rev. Lett.}, 1989, \textbf{63}, 980; \bibdoi{10.1103/PhysRevLett.63.980}.

\bibitem{16}  Rosenfeld Ya.,  J. Chem. Phys., 1993, \textbf{98}, 8126; \bibdoi{10.1063/1.464569}.

\bibitem{we} Wertheim M.S., J. Stat. Phys., 1986, {\bf 42}, 459; \bibdoi{10.1007/BF01127721}; \bibdoi{477.10.1007/BF01127722}.

\bibitem{13a} Ebeling W., Z. phys. Chem. (Leipzig), 1968, {\bf 238}, 400.

\bibitem{f2} Zhou Q., Yeh S., Stell G., J. Chem. Phys., 1995, {\bf 102}, 5785; \bibdoi{10.1063/1.469310}.

\bibitem{bor} Borowko M., Pizio O., Soko\l owski S. -- In: Computational Methods
in Surface and Colloid Science, Edited by Borowko M. New York, Marcel
Dekker, 2000, Chapter 4.

\bibitem{Fischer} Fischer J., Methfessel M., Phys. Rev. A, 1980, {\bf 22}, 2836; \bibdoi{10.1103/PhysRevA.22.2836}.



\bibitem{alejandre1} Gonz\'{a}lez-Melchor M., Alejandre J., Bresme F.,
Phys. Rev. Lett., 2003, {\bf 90}, 135506;
\\ \bibdoi{10.1103/PhysRevLett.90.135506}.

\bibitem{alejandre2} Gonz\'{a}lez-Melchor M., Alejandre J., Bresme F.,
J. Chem. Phys., 2005, {\bf 122}, 104710;
\\ \bibdoi{10.1063/1.1861878}.

\end{thebibliography}
\end{document}